\documentclass{emulateapj}

\usepackage{graphicx}
\usepackage{amsmath}
\usepackage{xspace}
\newcommand{\GRS}{GRS~1741.9-2853\xspace}
\newcommand{\nus}{{\em NuSTAR}\xspace}

\begin{document}

\lefthead{NuSTAR observation of GRS 1741.9-2853}
\righthead{Barri{\`e}re et al.}
\submitted{Accepted by ApJ}


\title{NuSTAR observation of a Type I X-ray burst from GRS~1741.9-2853}

\author{
Nicolas~M. Barri{\`e}re\altaffilmark{1},
Roman Krivonos\altaffilmark{1},
John~A. Tomsick\altaffilmark{1},
Matteo Bachetti\altaffilmark{2},
Steven~E. Boggs\altaffilmark{1},
Deepto Chakrabarty\altaffilmark{3},
Finn~E. Christensen\altaffilmark{4},
William~W. Craig\altaffilmark{1,5}, 
Charles~J. Hailey\altaffilmark{6},
Fiona~A. Harrison\altaffilmark{7}, 
Jaesub Hong\altaffilmark{8},
Kaya Mori\altaffilmark{6},
Daniel Stern\altaffilmark{9},
William~W. Zhang\altaffilmark{10}
}

\altaffiltext{1}{Space Sciences Laboratory, University of California, Berkeley, CA 94720, USA.}
\altaffiltext{2}{Institut de Recherche en Astrophysique et PlanŽtologie, UMR 5277, Toulouse, France.}
\altaffiltext{3}{MIT Kavli Institute for Astrophysics and Space Research, Massachusetts Institute of Technology, Cambridge, MA 02139, USA}
\altaffiltext{4}{National Space Institute, Technical University of Denmark, Copenhagen, Denmark.}
\altaffiltext{5}{Lawrence Livermore National Laboratory, Livermore, CA 94550, USA.}
\altaffiltext{6}{Columbia Astrophysics Laboratory, Columbia University, New York, NY 10027, USA.}
\altaffiltext{7}{Cahill Center for Astronomy and Astrophysics, Caltech, Pasadena, CA 91125, USA.}
\altaffiltext{8}{Harvard-Smithsonian Center for Astrophysics, Cambridge, MA 02138, USA.}
\altaffiltext{9}{Jet Propulsion Laboratory, California Institute of Technology, Pasadena, CA 91109, USA.}
\altaffiltext{10}{X-ray Astrophysics Laboratory, NASA Goddard Space Flight Center, Greenbelt, MD 20771, USA.}


\begin{abstract}
We report on two {\it NuSTAR} observations of GRS~1741.9-2853, a faint neutron star low mass X-ray binary  burster located 10$'$ away from the Galactic center. \nus detected the source serendipitously as it was emerging from quiescence: its luminosity was $6\times 10^{34}$ erg~s$^{-1}$ on 2013 July 31, and $5 \times 10^{35}$ erg~s$^{-1}$ in a second observation on 2013 August 3. A bright, 800-s long, H-triggered mixed H/He thermonuclear Type I burst  with mild photospheric radius expansion (PRE) was present during the second observation. Assuming that the luminosity during the PRE was at the Eddington level, a H mass fraction $X=0.7$ in the atmosphere, and a neutron star mass $M=1.4 M_{\odot}$, we determine a new lower limit on the distance for this source of $6.3 \pm 0.5$ kpc. Combining with previous upper limits, this places \GRS at a distance of 7 kpc. Energy independent (achromatic) variability is observed during the cooling of the neutron star, which could result from the disturbance of the inner accretion disk by the burst. The large dynamic range of this burst reveals a long power-law decay tail. We also detect, at a 95.6\% confidence level (1.7 $\sigma$), a narrow absorption line at $5.46\pm0.10$ keV during the PRE phase of the burst, reminiscent of the detection by \cite{waki.1984vn}. We propose that the line, if real, is formed in the wind above the photosphere of the neutron star by a resonant K$\alpha$ transition from H-like Cr gravitationally redshifted by a factor $1+z=1.09$, corresponding to a radius range of 29.0 -- 41.4 km for a mass range of 1.4 -- 2.0 $M_{\odot}$.
\end{abstract}




\section{Introduction}

\GRS (also called AX J1745.0-2855) is a very faint X-ray transient that was discovered by the {\it Granat} satellite in 1990 \citep{sunyaev.1990uq}. In 1996, {\it Beppo-SAX} detected a series of three Type I bursts from this source, establishing that it is a neutron star (NS) low mass X-ray binary (LMXB) \citep{cocchi.1999fk}. 
Since its discovery, \GRS has been detected in outburst several times by different observatories; in 1994, 1996, 2000, 2002, 2005, 2007, 2009, 2010, and in 2013 \citep{sakano.2002uq,muno.2003uq,wijnands.2005qf,wijnands.2006uq,wijnands.2007ve,muno.2007ly,porquet.2007ys, chenevez.2009zr,trap.2009vn,kennea.2009vn,degenaar.2010uq,degenaar.2010ys, degenaar.2013bh,kuulkers.2013dq}. Its typical outburst luminosity is $\sim$10$^{35-36}$ erg~s$^{-1}$ \citep[see ][for reviews of \GRS earlier and more recent outburst history]{trap.2009vn,degenaar.2014ve}. It has also been detected in quiescence by {\it Chandra}, with a 2--8 keV luminosity of  about 10$^{32}$ erg~s$^{-1}$ \citep{muno.2003uq}. Its outburst duty cycle is about 10\%, with typical outburst duration of $\sim$10 weeks and recurrence time of $\sim$ 2 years. Its average accretion rate in outburst is of the order of $10^{-10} \, M_{\odot}$~yr$^{-1}$ \citep{degenaar.2010ys}. 
A distance measurement  of 8 kpc was proposed by \citet{cocchi.1999fk}, placing the source very close to the Galactic center, as it lies $10'$ from Sagittarius A$^{\star}$. This distance was then further refined to an upper limit of  7~kpc by \citet{trap.2009vn}, who used {\it INTEGRAL}, {\it XMM-Newton} and {\it Swift} data of the 2005 and 2007 outbursts.

\GRS is a binary system where the companion is a low-mass star ($M \leq 1 M_{\odot}$) filling its Roche lobe and pouring matter onto an accretion disk. The  known LMXBs in our Galaxy are concentrated in the Galactic bulge \citep{grimm.2002vn,revnivtsev.2008kx}, and half of them are transients that increase their luminosity by several orders of magnitudes during outbursts \citep{liu.2007kx}. In such systems, X-ray outbursts are interpreted as an increase of the accretion rate onto the compact object \citep[e.g.][]{osaki.1974fk}. In the category of the {\it very faint} transients (2--10 keV luminosity in the $10^{34-36}$ erg~s$^{-1}$ range during outburst), which includes \GRS, the alternation between quiescence and outburst can be described by the thermal-viscous disk instability model, where the disk spends most of its time in a cold neutral phase (quiescence), and sometimes becomes hot and ionized, triggering accretion and causing an outburst \citep[see][for a review]{lasota.2001cr}. The source then goes back to quiescence, allowing the disk to refill until the cycle starts again. The very faint transients are not well studied, and in particular, the evolutionary scenario that leads to these very low time-averaged accretion rate systems is not clear \citep{king.2006fk}.

In many NS LMXBs outbursts are punctuated by Type I X-ray bursts, which appear as a sudden increase of luminosity (by up to two orders of magnitude) followed by an exponential or power-law decay with a duration ranging from a few seconds to several hundreds of seconds, and even up to a few tens of hours for the so-called super bursts. These bursts are thermonuclear flashes occurring at the surface of the NS \citep[see][for reviews]{lewin.1993zr,strohmayer.2006fk}. Their spectrum is well modeled by a black-body, arising from the hot photosphere of the NS during thermonuclear reactions and decaying during its subsequent cool down. These explosions result from the unstable burning of accreted H and/or He from the companion star. The accreted matter can fuse in different processes depending on its composition and the accretion rate \citep[e.g.][]{bildsten.1998fk,peng.2007fk}.


The observation of a Type I burst from an LMXB securely identifies it as a NS system. In addition, the study of these bursts probes both NS structure and fundamental physics. Their interpretation is rather simple in the first approximation: in the following we assume that the NS emits a pure Planck spectrum isotropically from its entire surface. Under this assumption, in the case a burst is bright enough to lift the photosphere ({\it i.e.} a burst with photospheric radius expansion, or PRE burst), the measurement of the luminosity (assumed to be the Eddington luminosity) and temperature can lead to the determination of the distance, and, in theory, of the radius and mass of the NS \citep{damen.1990fk}, which constrains the equation of state of ultra-dense matter \citep[e.g.][]{lattimer.2001vn,ozel.2006kx}. Determining the gravitational redshift at the surface of the NS (and thus the mass-to-radius ratio of the NS) can also be done via the observation of spectral features during the burst, either absorption lines \citep{van-paradijs.1979vn} or absorption edges \citep{int-zand.2010ly}. This is the most direct method of determining the mass-to-radius ratio, however detections of absorption lines are controversial \citep{waki.1984vn,nakamura.1988ys,magnier.1989zr,cottam.2002vn}.

In this paper we report on the detection of a Type I X-ray burst from the faint system \GRS; 
In section \ref{sec:obs} we describe the observations in which \GRS was detected, as well as the data analysis tools and methods that we used. The outburst and burst light curves are presented in Section \ref{sec:LC}. The different outburst phases are analyzed in section \ref{sec:outburst}, and the burst analysis is presented in section \ref{sec:burst}. An absorption line was found in the spectrum of the burst; it is described in section \ref{sec:absline}.  In section \ref{sec:oscillations}, we searched for a pulsation during the burst and the outburst but found none. Finally, a discussion of these results is presented in section \ref{sec:discussion}. Uncertainties at the 90\% confidence limit are quoted throughout this paper, unless noted.






\section{Observation, data reduction, and data analysis}
\label{sec:obs}
\nus is a hard X-ray (3--79 keV) focusing telescope with two identical, co-aligned telescopes producing a point spread function of $58''$ half power diameter ($18''$ full width at half maximum). Its effective area peaks at $\approx$900 cm$^2$ (adding up the two modules) around 10 keV, and its energy resolution at 10 keV is 400~eV \citep{harrison.2013ly}.

\GRS was detected serendipitously in two tiles of the Galactic survey \citep{harrison.2013ly} conducted by \nus on 2013 July 31 (obsID 40031001002, 44.4~ks on-time) and on 2013 August 3 (obsID 40031003002, 40.4~ks on-time). The source was brightening during the first observation as it was rising from quiescence \citep{degenaar.2014ve}. The outburst became more intense and underwent a bright Type I burst during the second observation, starting on 2013 August 3 at UT 23:03:15 (referred to as $t_s$ hereafter).

The data was processed by the \nus Data Analysis Software (NuSTARDAS) v1.3.0, which is distributed with HEASOFT v6.15, and uses the calibration files v20131007. Spectra and light curves were extracted using the ``nuproducts" FTOOL from the focal plane modules A and B (FPMA and FPMB). Our extraction region is centered on $\alpha_{\rm J2000} = 17^h45^m2.47^s$,  $\delta_{\rm J2000} = -28^{\circ}54'54.79''$, which is consistent within typical \nus astrometric uncertainty ($\leq 10''$) to the position reported in \citet{muno.2003uq}. A circular extraction region of $120''$ radius was used for the study of the burst, and a $60''$ radius was used for the study of the outburst. 

We performed the spectral analysis with the Interactive Spectral Interpretation System \cite[ISIS,][]{houck.2000fk}, using \citet{verner.1996kx} atomic cross sections  and \citet{wilms.2000vn}  abundances  for the interstellar absorption. We used $\chi^2$ statistics to evaluate the quality of the fit of the different spectral models. We analyzed  the data over the range 3--30~keV as both the outburst and the burst spectra cut off sharply  above 30~keV. The spectral binning is identical for each spectrum presented in this paper: spectra of FPMA and FPMB were combined, and the spectral bins were grouped to reach at least 50 source $+$ background counts and 3 $\sigma$ signal-to-noise significance in each group. The study of the outburst uses a different region from the same detector chip as background (identical time), and the study of the burst uses the same region as used for source extraction, but taking times preceding the burst as background.

\section{Light curve analysis}
\label{sec:LC}

\begin{figure}[t]
\begin{center}
\includegraphics[width=0.47\textwidth]{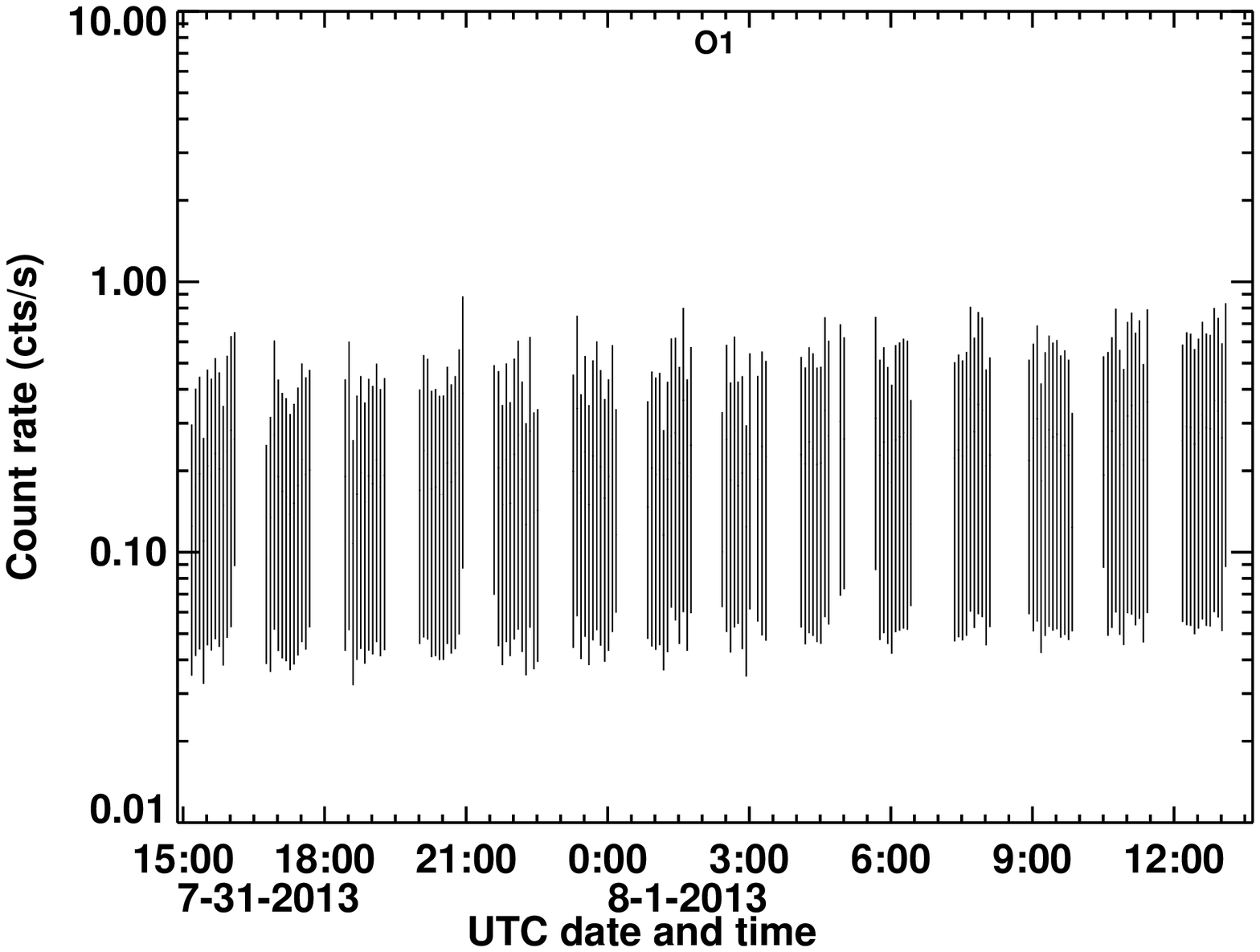}
\includegraphics[width=0.47\textwidth]{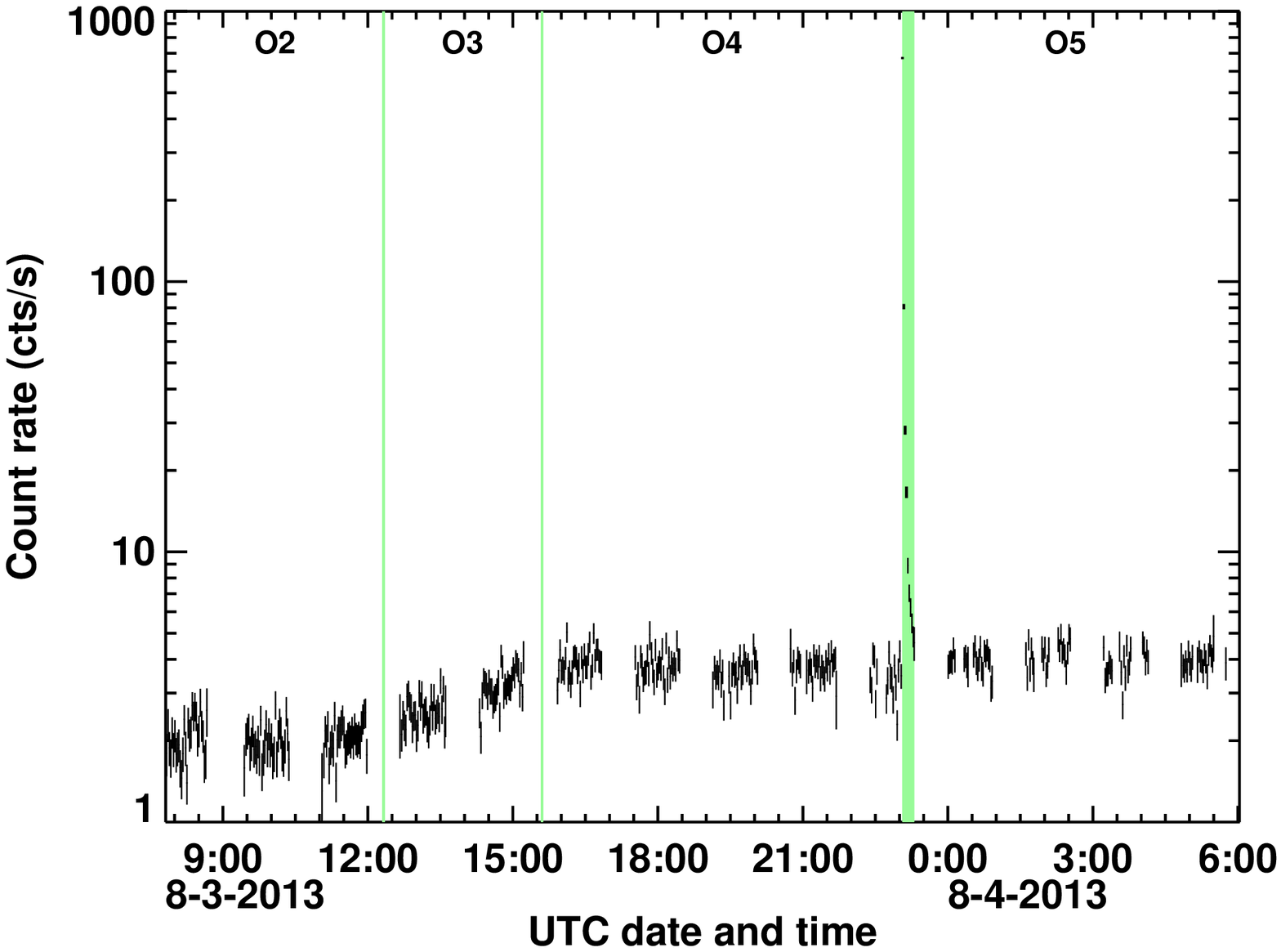}
\caption{Light curve of \GRS  during obsIDs 40031001002 (300-s bins, left panel) and 40031003002 (100-s bins, right panel) in the 3--30 keV band. Instrumental background is subtracted and FPMA and FPMB are combined. Error bars show the 1-$\sigma$ uncertainty level. For spectral analysis, the outburst is split into five phases as shown by the green vertical lines and marked by the labels at the top of the plot frames (the first phase, O1, is simply defined as the first observation). 
\label{fig:fullLC}}
\end{center}
\end{figure}

The light curves were corrected for dead time (which reaches 45\% at the peak of the burst), finite extraction region, and vignetting effects. Figure \ref{fig:fullLC} shows the 3--30 keV light curves of the two observations. The first observation (hereafter referred to as O1) shows a continuous flux rise, revealing the beginning of the outburst \citep{degenaar.2014ve}. In the second observation (right panel), five periods are distinguished: the ``low level" outburst at the beginning of the observation (O2), the rise (O3), the pre-burst (O4), the burst, and post-burst (O5). 

\begin{figure}[t]
\begin{center}
\includegraphics[width=0.47\textwidth]{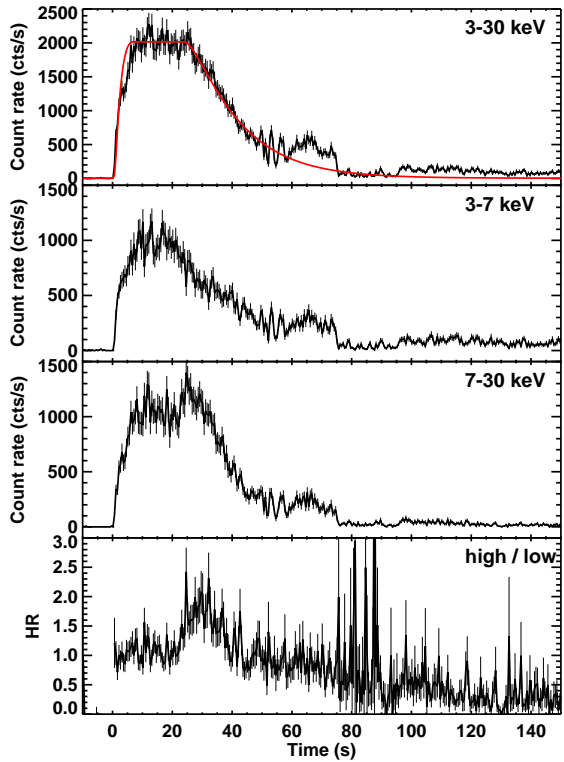}
\caption{Burst light curve (0.5-s bins) in the 3--30 keV, 3--7 keV and 7--30 keV bands, and ratio of the 7--30 keV to the 3--30 keV bands. The origin of the time marks the beginning of the burst at 23:03:15 UT on 2013 August 3. In the top panel, the red curve shows the best fit using a modified FRED curve to accommodate a plateau at the peak.
\label{fig:burstLC+fit}}
\end{center}
\end{figure}

Figure \ref{fig:burstLC+fit} shows a close-up view of the burst with finer bins. The burst rise, peak, and early decay is well described by a Fast Rise Exponential Decay (FRED) function, with a plateau inserted at the peak: 
\begin{align}
f(t) = \left\{
\begin{array} {l l}
 \exp \left( -{\tau_R \over t} - {t \over \tau_D} \right)  & 0 < t \leq t_{\rm peak} \\
 \exp \left( -2 \sqrt{ \tau_R \over \tau_D} \right) & t_{\rm peak} < t \leq  t_{\rm peak} + t_{\rm plateau} \\
 \exp \left( -{\tau_R \over t - t_{\rm plateau}} - {t - t_{\rm plateau} \over \tau_D} \right)  & t > t_{\rm peak} + t_{\rm plateau} \\
\end{array}
\right.
\end{align}
with, $t_{\rm peak} = \sqrt{\tau_R \tau_D}$, the time to reach the peak luminosity,  $\tau_R$ and $\tau_D$ the characteristic rise and decay times, respectively, and $t_{\rm plateau}$ the duration of the plateau. Binning the light curve to 0.5-s bins, the best fit to the data returns a rise time of 4.2~s (time to reach 90\% of the peak value), $t_{\rm plateau}= 18$~s, and an e-folding time of 20.6~s (time to decrease by a factor $e^1\approx 2.72$). The burst lasts over 800~s (\nus did not catch the end of the decay because of an Earth occultation), and its
decay is unusual with variability between 48~s to 95~s into the burst. The hardness ratio remains constant within errors during these variations, which is similar to the energy-independent ({\it i.e.} achromatic) late-time variability observed by \citet{int-zand.2011uq}. 

\begin{figure}[t]
\begin{center}
\includegraphics[width=0.47\textwidth]{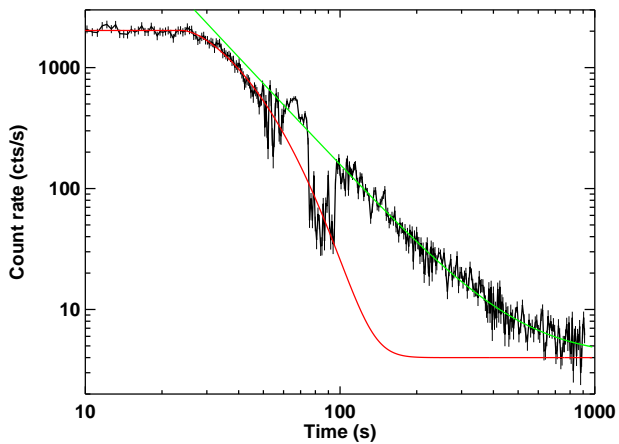}
\caption{Burst decay light curve in the 3--30 keV band, with 0.5~s, 1~s, 2~s, 4~s, and 8~s bins, in the 0--60~s, 60--120~s, 120--240~s, 240--480~s, and 480-1000~s ranges, respectively. The red curve shows the modified FRED curve fit to the 0--50~s interval of the light curve (identical to the model presented in Fig. \ref{fig:burstLC+fit}), and the green curve shows a power-law fit to the 110--900~s interval. Both models include a constant equal to 4 cts/s to account for the persistent emission.
\label{fig:burstdecay}}
\end{center}
\end{figure}

At 95~s, the flux increases by a factor of 3--4 and is then followed by a smooth decay that is not well fit by an exponential curve. Figure \ref{fig:burstdecay} shows that, instead, it is better described by a power-law, as expected in burst tails when nuclear reactions no longer take place \citep[][and references therein]{int-zand.2014uq}. Fitting to the burst tail (110--900~s) a function defined as:
\begin{equation}
f(t) = C_0 \left( {t-t_s \over t_0-t_s} \right)^{-\alpha}
\end{equation}
and setting the time of the first point in the fit $t_0 = 110$~s ($C_0$ is the count rate at $t_0$, and $t_s$ is the time of the burst onset), we find a decay index $\alpha=2.25 \pm 0.03$. As noticed by \citet{int-zand.2014uq}, $\alpha$ is strongly correlated with $t_s$, however we estimate that the error on $t_s$ cannot exceed $\pm 0.2$~s, which leads to a systematic error on $\alpha$ lower than $\pm$0.002. Despite some variability, the interval from 75~s to 95~s seems to match with the extrapolation of the exponential decay. Another noticeable fact is that the two cooling curves do not intersect. These results are discussed in section \ref{sec:discussion}.


\section{Outburst}
\label{sec:outburst}

\begin{figure}[t]
\begin{center}
\includegraphics[width=0.48\textwidth]{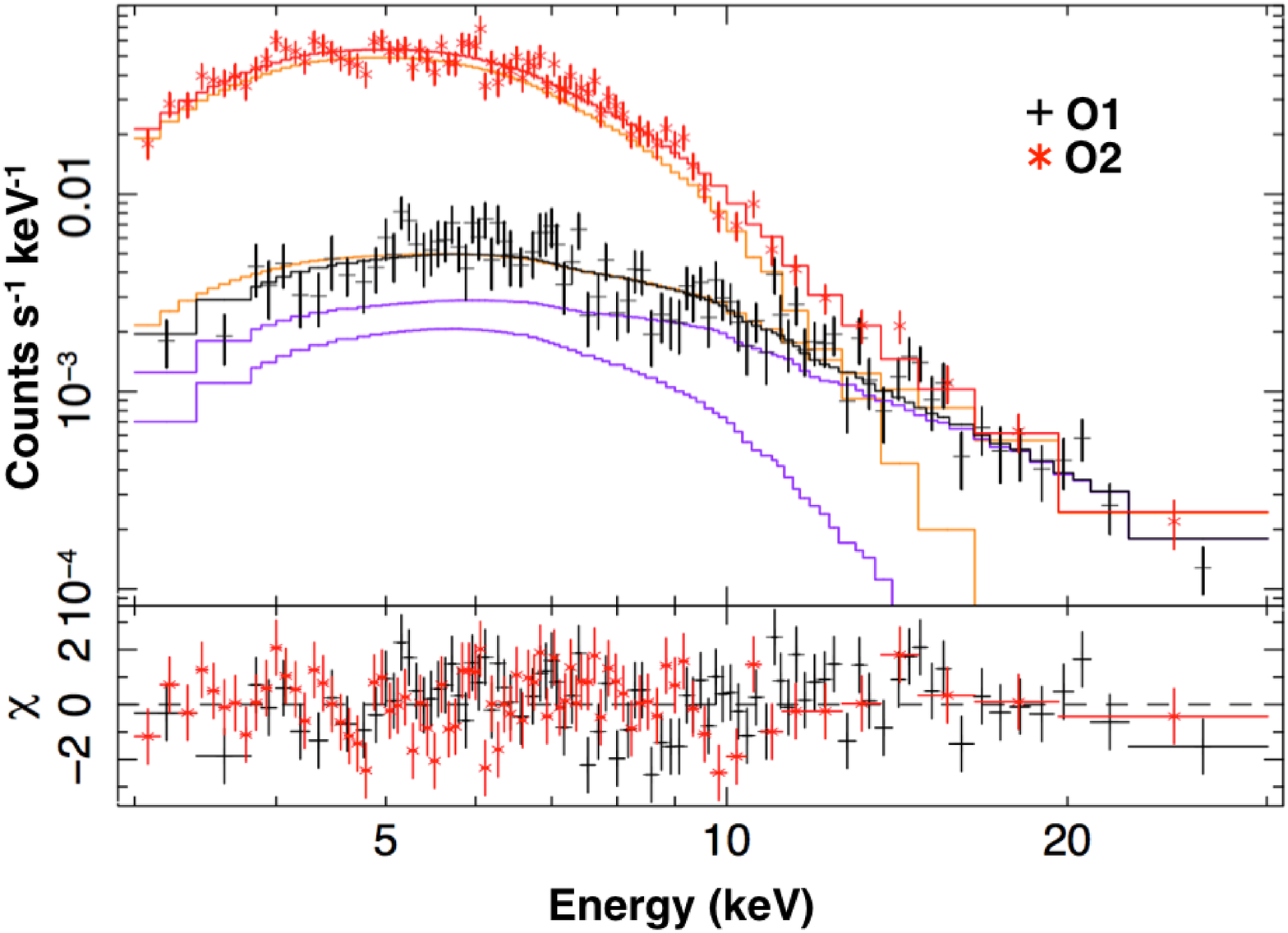}
\includegraphics[width=0.48\textwidth]{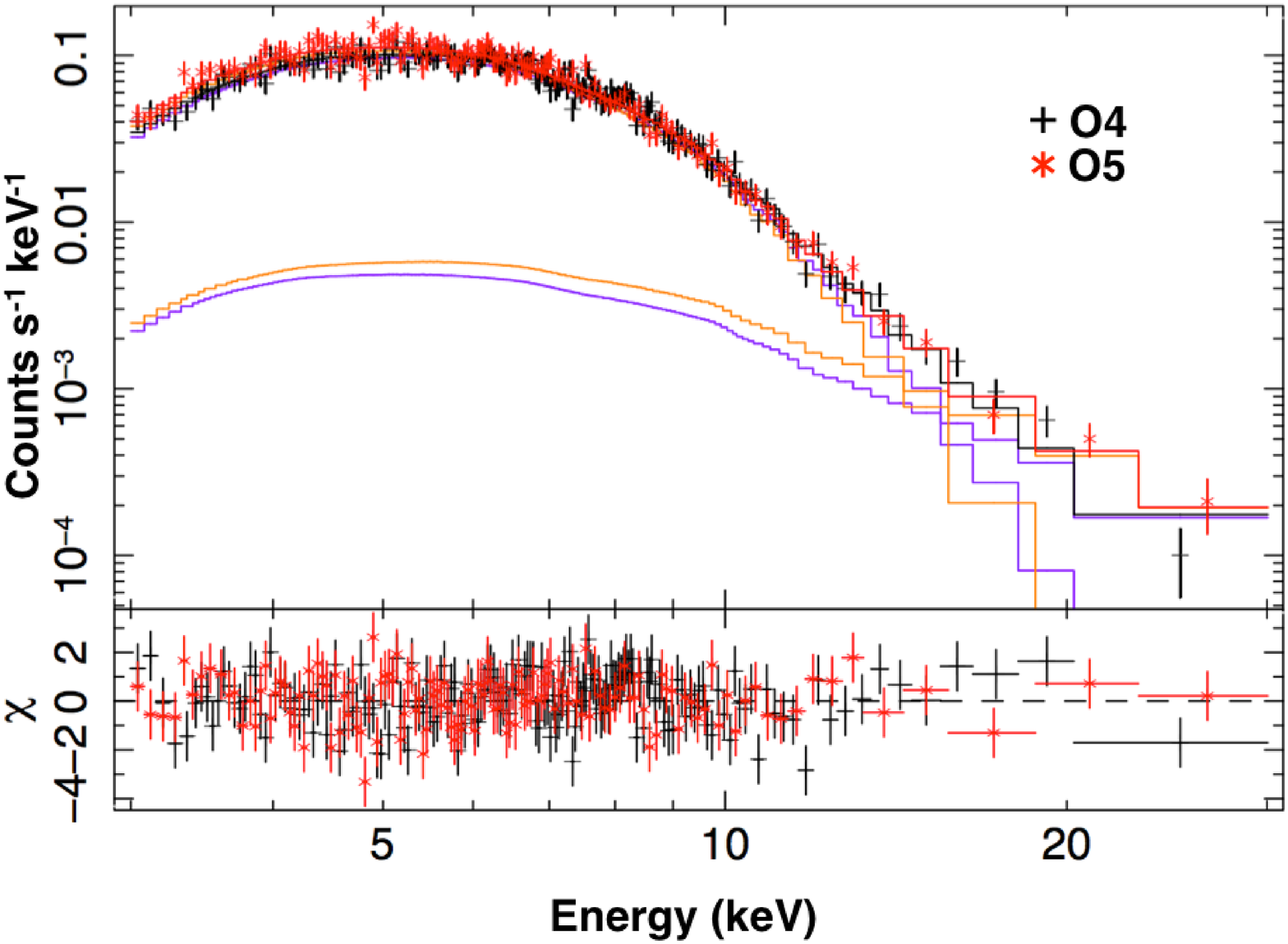}
\caption{Spectra of the outburst phases with their best fit model and the decomposition of the model into the black-body component and the power-law component (high-energy tails). The left panel shows the spectrum of phases O1  (black plus signs, and model breakdown in purple), and O2 (red crosses, and model breakdown in orange). The right panel shows phases O4 (black plus signs, and model breakdown in purple) and O5 (red crosses, and model breakdown in orange). Each spectrum is the combination of data from modules A and B. The data points are shown with 1-$\sigma$ error bars, and the solid lines show the best fit model. The lower panels show the deviation from the model in units of standard deviation.
\label{fig:outburstspc}}
\end{center}
\end{figure}

We fit the four different phases of outburst identified in section \ref{sec:LC} (first observation, and low level, pre-burst and post-burst in the second observation: O1, O2, O4, O5, respectively) with an absorbed black-body + power-law model ($\texttt{TBabs} \times( \texttt{bbody} + \texttt{powerlaw}))$ with tied column density. This model is physically interpreted as the thermal emission of the disk and the NS surface (in the present case, these two components are indistinguishable and are thus modeled with a single black body) in addition to the non-thermal emission from a corona (Compton scattering of thermal photons by hot electrons). We obtain good agreement between the model and the data with a $\chi^2_\nu=1.15$ (461 dof), yielding a column density $N_H = 8.3^{+1.9}_{-1.3} \times 10^{22}$~cm$^{-2}$ (Figure \ref{fig:outburstspc}). The black-body temperatures and power-law photon indices are presented in Table \ref{tab:accrate}, along with the corresponding bolometric fluxes, luminosities, and derived accretion rates.

The column density that we find is low compared to previous studies. For instance, \citet{muno.2003uq} found $N_H = (9.7 \pm 0.2) \times 10^{22}$ cm$^{-2}$ using 2--7 keV data from {\it Chandra}, \citet{trap.2009vn} found $N_H = (11.9 \pm 0.2) \times 10^{22}$ cm$^{-2}$ using 2--8 keV data from {\it XMM-Newton}, and \citet{degenaar.2010ys} found $N_H = (14.0 \pm 0.7) \times 10^{22}$ cm$^{-2}$ using 2--10 keV data from {\it Swift}/XRT. In these works, the results were obtained using the default XSPEC cross-sections and abundances, which leads to $N_H$ values $\approx$ 33\% lower than those obtained with $\texttt{TBabs}$ using \citet{wilms.2000vn} abundances and \citet{verner.1996kx} cross-sections \citep{nowak.2012zr}. However, we note that these past studies were done over smaller energy ranges and used a simple absorbed power-law model, which requires a higher column density than a black-body to get a curved spectrum at low energies.

The strong gravitational field of the NS decreases the luminosity observed by a distant observer (noted with the subscript $\infty$) with respect to the one emitted in the local frame (subscript $\star$) by a factor $(1+z)^2$, where $1 + z = \left(1-2GM / (R_\star c^2) \right)^{-1/2}$. In this equation, $c$ is the speed of light, $M$ and $R$ are the mass and radius of the NS, and $G$ is the gravitational constant. It also affects the emitted spectrum by changing the temperature observed by a distant observer $T_{\infty} = T_{\star} (1+z)^{-1}$ \citep{lewin.1993zr}. Assuming the canonical values $M = 1.4 M_{\odot}$ and $R_\star = 10$~km, one finds $1 + z \approx 1.31$. Unless specifically noted with the subscript $\star$, all physical quantities will hereafter be expressed as seen by a distant observer.

The unabsorbed bolometric luminosity can be converted into an accretion rate per unit area $\dot{m}$ using the formula\footnote{ This equation assumes that the total luminosity of the system is powered by accretion (i.e. emission from the accretion disk, the corona and the NS surface), and that obscuration is negligible. This is equivalent to saying that the radiation efficiency is equal to 1.} $L_b  =  4\pi R_{\star}^2 \, \dot{m} \,(G M / R_{\star})(1+z)^{-1}$ \citep{galloway.2008fk}, which can be inverted to show:
\begin{equation}
\begin{split}
\dot{m}  = 32.8 \times
   \left( \frac{F_{b}}{10^{-11} {\rm erg \, cm^{-2}\, s^{-1}}}\right)
   \left( \frac{M}{1.4 \, M_{\odot}}\right)^{-1}
   \left(\frac{1 + z}{  1.31}\right)  \\
   \times \left(\frac{R_{\star}}{10 \, {\rm km}}\right)^{-1}
    \left( \frac{d}{ 7 \,{\rm kpc}}\right)^2
   \; {\rm g \, cm^{-2} \,s^{-1}}
\end{split}
\label{eq:accretionrate}
\end{equation}
where $d$ is the distance to the observer and $F_{b}$ the bolometric flux seen by a distant observer.

\begin{deluxetable*}{ccccccccc}
\tabletypesize{\footnotesize}
\tablecaption{Outburst fit parameters, flux, luminosity (assuming d=7.0 kpc), and derived accretion rate per unit area at the NS surface for the different phases identified in Section~\ref{sec:LC}.  \label{tab:accrate}}
\tablecolumns{9}
\tablewidth{0pt}
\tablehead{
\colhead{Outburst} 	& \colhead{$k_B T$ (keV)} & \colhead{$\Gamma$} & \colhead{$F_{2-10}$ \tablenotemark{a} }	& \colhead{$F_{b}$ \tablenotemark{b} } 	& \colhead{$L_{b}$ \tablenotemark{c} } & \colhead{$\dot{m} $ \tablenotemark{d}} 	& \colhead{$\dot{m}_{\rm Edd} $ \tablenotemark{e}} 	& \colhead{$\dot{M} $ \tablenotemark{f}} 	\\
 phase 		     	&  			& 					&  	&   	&	&  &	}
\startdata
O1 	& $1.74^{+0.66}_{-0.30}$ 	& $1.27^{+0.31}_{-0.40}$ & $0.184^{+0.015}_{-0.014}$ 	& $0.987^{+0.144}_{-0.145}$ 	&  $0.579^{+0.084}_{-0.085}$	&  32.4  	& $3.6 \times 10^{-4}$& 0.64  	\\
O2 	& $1.35\pm0.05$ 		& $1.50^{+0.73}_{-0.97}$ & $3.01^{+0.13}_{-0.12}$ 		& $ 5.21^{+0.42}_{-0.40}$ 		& $3.06^{+0.25}_{-0.23} $ 		&  171 	& $1.9 \times 10^{-3}$& 3.4	\\
O4  	& $1.46\pm0.03$ 		& $1.69^{+0.72}_{-0.97}$ & $5.58\pm0.18$ 			& $8.04^{+0.38}_{-0.35}$ 		& $4.71^{+0.22}_{-0.21} $ 		&  264 	& $3.0 \times 10^{-3}$& 5.3	\\
O5 	& $1.39\pm0.04$ 		& $1.57^{+0.98}_{-1.49}$ & $6.16^{+0.22}_{-0.21}$ 		& $9.14^{+0.52}_{-0.48}$ 		& $5.36^{+0.30}_{-0.28} $ 		&  300 	& $3.4 \times 10^{-3}$& 6.0
\enddata
\tablenotetext{a}{Unabsorbed 2--10 keV flux in units of $10^{-11}$ erg cm$^{-2}$ s$^{-1}$.}
\tablenotetext{b}{Unabsorbed bolometric flux (evaluated over the 0.1--100 keV energy range) in units of $10^{-11}$ erg cm$^{-2}$ s$^{-1}$.}
\tablenotetext{c}{Unabsorbed bolometric luminosity assuming a distance of 7 kpc in units of $10^{35}$ erg~s$^{-1}$.}
\tablenotetext{d}{Accretion rate per unit area in g cm$^{-2}$ \,s$^{-1}$ determined with equation \ref{eq:accretionrate}.}
\tablenotetext{e}{Accretion rate per unit area normalized by the Eddington accretion rate  $\dot{m}_{\rm Edd} = 8.8 \times 10^4$ g cm$^{-2}$ \,s$^{-1}$ (assuming $M = 1.4 M_{\odot}$ and $R = 10$ km). }
\tablenotetext{f}{Global accretion rate in units of $10^{-11}$ M$_{\odot}$  yr$^{-1}$, assuming a NS radius of 10 km.}

\end{deluxetable*}

\section{Type I burst}
\label{sec:burst}

\begin{figure}[t]
\begin{center}
\includegraphics[width=0.46\textwidth]{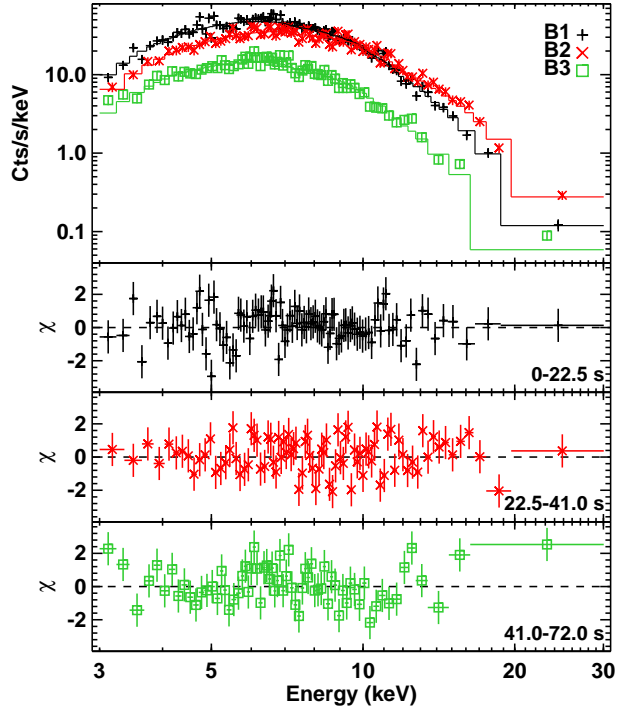}
\caption{Spectrum of the burst recorded during three time intervals B1 (0 -- 22.5 s, black plus signs), B2 (22.5 -- 41.0~s, red crosses), and B3 (41.0 -- 72.0~s, green squares). Each point is the combination of FPMA and FPMB, and the errors are 1-$\sigma$. The lower panels show the residuals between the best-fit models (absorbed black-body) and the data in units of standard deviation.
\label{fig:burstspc}}
\end{center}
\end{figure}

We first look at the burst spectrum divided into 3 phases, corresponding to the PRE (0~--~22.5~s,  B1), the high-temperature (after the photosphere falls back onto the NS surface, commonly referred to as {\it touch-down}, 22.5~--~41.0~s, B2), and the beginning of the decay (41.0~--~72.0~s, B3), as defined in the next paragraph. We use the O3 outburst phase (14.6~ks right before the burst) as background. The cross calibration between FPMA and FPMB is excellent, we found agreement to better than 0.2\%, so we removed the normalization constant from the model. The three spectra are well fit with an absorbed black-body (we use $\texttt{TBabs} \times \texttt{bbody}$) with tied column density ($\chi^2_{\nu}$ = 1.06, 255 dof, Figure \ref{fig:burstspc}). We find $N_H = (11.3 \pm 1.5) \times 10^{22}$~cm$^{-2}$, consistent within errors with the value found for the outburst, and black-body temperatures of $2.11 \pm 0.04$~keV, $2.65 \pm 0.06$~keV and $1.97 \pm 0.05$~keV for spectra B1, B2 and B3, respectively.

\begin{figure}[p]
\begin{center}
\includegraphics[width=0.49\textwidth]{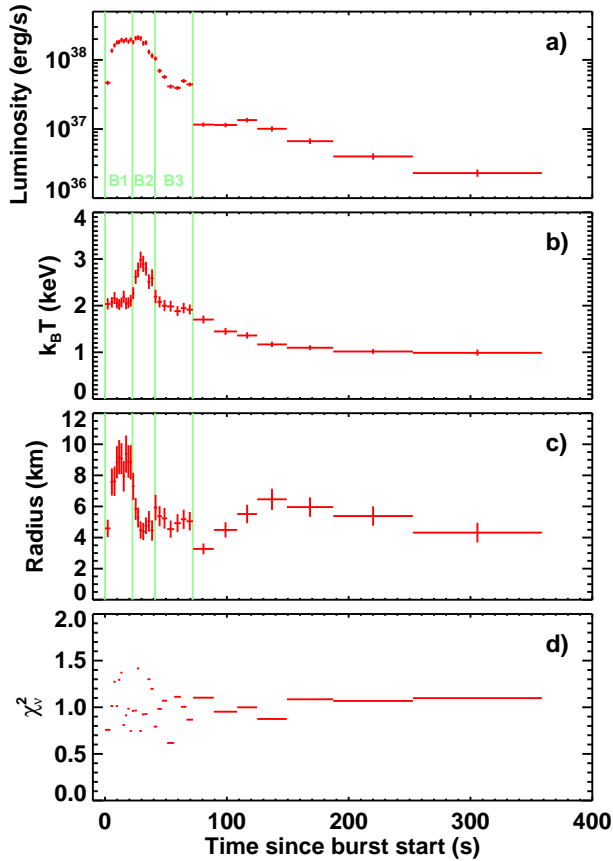}
\caption{Time-resolved black-body spectral fit of the burst, assuming a distance of 7 kpc to the source, and keeping the column density fixed to $1.13 \times 10^{23}$~cm$^{-2}$ . Panel a) shows the bolometric black-body luminosity. Panel b) shows the black-body temperature, and  panel c) shows the neutron star photospheric radius inferred from the black-body flux and temperature. Panel d) shows the reduced $\chi^2$ of each fit. All quantities are as seen by a distant observer. The green vertical lines define the three time intervals B1, B2 and B3 used in Figure \ref{fig:burstspc}.
\label{fig:burst3panel}}
\end{center}
\end{figure}

Next we look at the time-resolved spectral properties of the burst. The burst is divided to yield 700 net counts (FPMA + FPMB) in each time slice, from $t_s$ to $t_s+359$ s. The top panel of Figure \ref{fig:burst3panel} shows the luminosity derived from the black-body fit to the time-resolved spectra, assuming a distance to the source of 7~kpc. The middle panel of Figure \ref{fig:burst3panel} shows the black-body temperature and the bottom panel shows the neutron star photospheric radius $R$ inferred from the black-body luminosity $L$ and temperature $T$:
\begin{equation}
R = \sqrt{L \over {4\pi \, \sigma T^4}},
\label{eq:BBradius}
\end{equation}
with $\sigma$ being the Stephan-Boltzmann constant.

Integrating the luminosity over these 32 slices, we find that the total energy radiated is $9.5 \times 10^{39}$ erg (at 7 kpc). Following \citet{galloway.2008fk}, we calculate the burst characteristic time $\tau = E_b / F_{pk}$, where $E_b$ is the fluence of the burst (i.e. the integrated bolometric burst flux) and $F_{pk}$ is the peak bolometric flux. With $E_b = 1.61 \times 10^{-6}$ erg cm$^{-2}$ and $F_{pk} = 3.58 \times 10^{-8}$ erg cm$^{-2}$ s$^{-1}$, we find $\tau=45.0$ s. We also note the very large dynamic range of this burst, which appeared on top of weak persistent emission; the ratio of the burst peak bolometric flux to the persistent bolometric flux preceding the burst is 445, amongst the highest ever observed \citep{int-zand.2014uq}. 

As pointed out by \citet{worpel.2013vn}, the persistent flux can vary during bursts, which limits the accuracy of our method. We thus tried to apply their analysis to this dataset. We subtracted the instrumental background from the burst time-resolved spectra, and fitted them with the model {\tt TBabs} $\times$ ({\tt blackbody} $+ f_a  \times$ PE), PE being the best fit model to the persistent emission preceding the burst with all parameters fixed, and $f_a$ being a constant to account for possible variability of the PE. The maximum value of $f_a$ is found during the PRE phase of the burst, however it is very poorly constrained: $f_a = 122^{+154}_{-122}$. This is explained by the fact that the ratio between  the PRE emission ($\approx L_{\tt Edd}$) and the persistent emission ($3\times10^{-3} L_{\tt Edd}$)  is about 333, while the two spectra have fairly similar shapes allowing the persistent spectrum to ``nest'' into the burst spectrum without any constraint on $f_a$. We thus are not able to constrain the PE variability during the burst.

\subsection{Presence of an absorption line}
\label{sec:absline}

\begin{figure}[t]
\begin{center}
\includegraphics[width=0.5\textwidth]{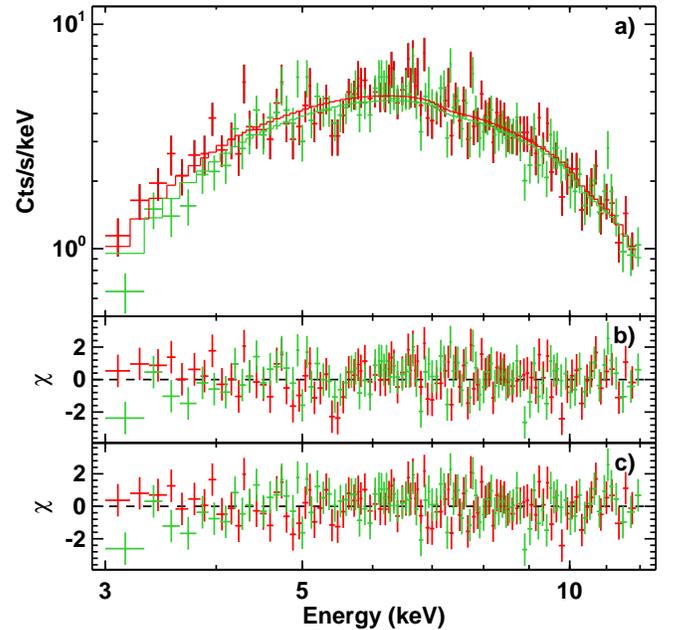}
\caption{FPMA (red) and FPMB (green) spectra of the burst during time interval B1(the PRE phase), with 1-$\sigma$ error bars. 
Panel a) shows the data with the absorbed blackbody best fit model, and panel b) shows the residuals for this model in units of standard deviation. Panel c) shows the residuals when a Gaussian absorption line at 5.46 keV is added to the model. 
\label{fig:absline}}
\end{center}
\end{figure}

We observe the presence of a weak transient spectral feature during the PRE, shown in Figure \ref{fig:absline}. We tested an absorption edge to model the feature but it did not fit, returning a null depth. The best fit is obtained using a narrow (poorly constrained width) Gaussian absorption line centered at $5.46 \pm 0.10$ keV, which only marginally improves the $\chi^2_{\nu}$ from 1.02 (216 dof) to 0.99 (213 dof), keeping $N_H$ fixed to $1.13 \times 10^{23}$ cm$^{-2}$ in both cases. The Gaussian normalization parameter is $-0.029\pm 0.016$ ph~cm$^{-2}$~s$^{-1}$, yielding an equivalent width of $0.08\pm 0.06$ keV.

We first performed a sanity check using a much smaller region of $60''$ radius, which makes it less sensitive to background variations. The line was still visible with this reduced extraction region. Next, we determine the significance using the XSPEC \citep{arnaud.1996uq} script {\tt simftest}\footnote{simftest is a script that generates fake spectra based on the model {\it without} the component under investigation, and then fit to them the model with and without the component. The difference in fit statistics is recorded for each trial, and is then compared to the difference obtained with the actual dataset. If a significant number of trials return a difference in fit statistics larger than that obtained with the real dataset, we conclude that the component under investigation is not required by the data, as the spectral feature likely results from a statistical fluctuation.} 
with $5,000$ trials (constraining the line centroid in the 5.26--5.66 keV range, and the line width to be $< 0.05$ keV), we found that the probability for the absorption line to be required by the data is 98.5\% (equivalent to 2.4 $\sigma$, as it is a one-sided distribution). This test was performed with the model $\texttt{constant} \times \texttt{TBabs} \times ( \texttt{BBody} + \texttt{Gauss})$ jointly fit to FPMA and FPMB spectra over the 3--15 keV range, with $N_H$ fixed to $1.13\times 10^{23}$~cm$^{-2}$. The spectra are rebinned to at least 25 counts and 3~$\sigma$ per bin (the former condition insures a bin significance higher than 5~$\sigma$ in the 5--6 keV range), as shown in Figure \ref{fig:absline}.
Accounting for the fact that we found the line in 1 of 3 sub-spectra (spectra B1, B2, B3), the significance of the line drops to 95.6\% (1.7 $\sigma$).

\subsection{Search for oscillations}
\label{sec:oscillations}

In April and July 1996, \citet{strohmayer.1997kx} reported strong detection of oscillations at a frequency of 589 Hz during three bursts observed with {\it RXTE} coming from a direction close to \GRS.  The strongest oscillations were detected during the brightest of the three bursts with a fraction\footnote{defined as the amplitude of the sinusoid divided by the constant count rate below the oscillations} of $(8.4 \pm 1.4)$\% during the peak and $(18.4 \pm 2.8)$\% during the decay, and were seen only above 8 keV. These bursts were tentatively attributed to MXB 1743--29 (a LMXB $24'$ away from \GRS), with the 99\% confidence region excluding the position of \GRS. In August-September 1996, \citet{cocchi.1999fk} observed the Galactic center with {\it Beppo-SAX} and detected three bursts from \GRS, determining that it is a transient LMXB. \citet{galloway.2008fk} made the association between \GRS and the oscillations detected by \citet{strohmayer.1997kx} because these bursting episodes happened only a few months apart, and had similar characteristics.

We ran a combination of timing analysis procedures in order to look for coherent pulsations in the \nus dataset. One difficulty with \nus timing is dead time: each event is followed by a dead time of $\approx$2.5\,ms \citep{harrison.2013ly}. At high count rates, such as those reached during the burst in our observation, dead time affects the frequency response of the focal plane modules, producing a ``wavy'' pattern in the power density spectrum (PDS; \citealt{VDK89,Zhang+95}), which affects heavily our detection limits for pulsations. 

We first analyzed the burst light curve with a running-window epoch folding \citep{Leahy+83fold} around 589 Hz. We used a number of different windows (over the range 2--400\,s) and calculated the $\chi^2$ of the putative pulse profiles. We found no signals exceeding the 90\% detection level, taking into account the number of trial periods per interval and the number of time intervals that the burst was divided into \citep{Leahy+83fold}. We then searched for pulsations at any frequency during the burst. To work around the distortions of the white noise level of the PDS due to dead time, we followed \citet{Bachetti+14} and used the co-spectrum as a proxy for the PDS: the Fourier transform of the signal from FPMA is multiplied by the conjugate of the signal from FPMB, producing the cross power density spectrum (CDPS), which contains in its real part (or co-spectrum) the signal that is correlated between the two FPMs. Since the per-event dead time is uncorrelated between the two detectors, the co-spectrum does not show the deformations of the white noise level that affect the PDS.

We analyzed the {\em dynamical} co-spectrum during the burst, looking for transient peaks that might indicate a pulsation over intervals of 2--100\,s in a sliding window. The burst, and in particular its rising and decay phases, were scrutinized, as well as the entire observation covering the different outburst phases. We searched for pulsations in the 3--30 keV and 8--30 keV bands. We did not find any significant power at any frequency (following the method of \citet{VDK89}, and taking into account the prescription in \citet{Bachetti+14}), which means that we did not detect any pulsation in the burst nor in the outburst phases. The only variability found during the burst was a strong increase of the red noise level, as it is expected from burst light curves \citep[e.g.][]{TerrelOlsen70}. 

To estimate upper limits, we simulated a large number of event lists containing a sinusoidal pulsation, with the observed count rates and durations for each interval (B1, B2, B3, burst tail), and estimated the pulsed fraction for which the signal was detected 90\% of the time above the \citet{Leahy+83fold} detection level for the relevant technique (epoch folding, assuming $n_{trials}=1000$ or PDS/cospectrum). We find that our upper limits at 589~Hz, calculated with epoch folding are always comparable or above the pulsed fraction measured by \citet{strohmayer.1997kx} ($\sim20$\% in B1 and B2, going to 25\% in B3 and in the burst tail), and therefore we cannot rule out the presence of these oscillations in our data set. Regarding the search at all frequencies, the upper limit depends on the frequency, particularly because of the suppression of low frequency modulation in dead time affected datasets \citep[see][]{Bachetti+14}. At high frequencies ($\sim600$~Hz), the upper limits are similar to those determined with epoch folding, at low frequencies ($\sim50$~Hz), the upper limit is higher in the high-count rate intervals, 30--40 \% for B1 and B2.

\section{Discussion}
\label{sec:discussion}

Figure \ref{fig:burst3panel} shows indications of photospheric radius expansion (PRE) during the first 20 s of the burst, where the temperature is constant at $\approx 2.1$ keV. The sudden increase in temperature that follows is interpreted as the ``touch down'', when the photosphere falls back onto the NS. This is far from being an extreme PRE event, as the X-ray flux remains measurable during the expansion phase (the adiabatic cooling can be so extreme in the case of super-expansion bursts that the bulk of the flux moves into the UV band), and the apparent black-body radius\footnote{not corrected for the relativistic effects} does not exceed 10 km (\ref{fig:burst3panel}), but the constant bolometric flux and the black-body temperature evolution (which determine the apparent black-body radius, see eq. \ref{eq:BBradius}) are clearly suggestive of a PRE.

The value of $\tau=45.0$~s that we determined in section \ref{sec:burst} puts this burst in the category of the long and unusual ones for this source, most of its ``regular'' bursts having $\tau \sim 10$s \citep{galloway.2008fk}. This is not to be confused with super-bursts lasting longer than 1~h, which results from the slow burning of C \citep[e.g.][]{strohmayer.2002vn}, nor with intermediate bursts lasting $\sim 30$ min, thought to arise from the burning of a thick layer of pure He \citep[e.g.][]{int-zand.2005zr}.

\GRS is known to produce both pure He and mixed H/He bursts \citep{trap.2009vn}. The accretion rate increased during our two observations to reach 0.003~$\dot{m}_{\rm Edd}$ before the burst, well below the critical value $\dot{m}_C = 0.01\, \dot{m}_{\rm Edd}$ for which there is no stable burning of the freshly accreted H, nor of the He, assuming solar CNO abundance ($Z_{CNO}$) \citep[][]{fujimoto.1981uq,bildsten.1998fk,narayan.2003fk}. We note that $\dot{m}_C$ is proportional to ${Z_{CNO}}^{1/2}$, i.e. only weakly dependent on our assumption of solar abundance \citep[see equation 36 in][]{bildsten.1998fk}. \citet{peng.2007fk} show that once a sufficient column density is reached, unstable H burning can trigger the burning of a thick He layer that formed by sedimentation for $\dot{m} \lesssim 0.003\, \dot{m}_{\rm Edd}$, resulting in a powerful long burst. This matches well with what we observed; Hydrogen burning is $\beta$-limited, which slows down the reaction compared to helium burning, hence creating a longer burst. However, the burst would have been much weaker if only H had burned.
Thus, based on the relatively long duration of the burst and on the low accretion rate preceding the burst, we conclude that we observed an H-triggered, mixed H/He, burst.

The conclusion that this burst was H-rich is now used to calculate the corresponding Eddington luminosity, allowing us to determine a distance for \GRS. In the case of a PRE burst, the peak flux $F_{pk}$ corresponds to the Eddington luminosity $L_{Edd}$, and the distance $d$ of the object is simply given by
\[
d = \left( {L_{Edd}} \over{4\pi\, F_{pk}}  \right)^{1/2}.
\]
The Eddington luminosity, as seen by a distant observer, is given by \citep{lewin.1993zr}:
\[
L_{Edd} = \left( {4 \pi c G M_{NS}} \over {\kappa} \right) \left( {1+z} \right)^{-1},
\]
where $\kappa$ is the electron scattering opacity during the expansion phase. In the case where the scattering electrons become relativistic, the opacity can be approximated by $\kappa = 0.2\times(1+X)\,\left[ 1+ (kT/39.2 \,{\rm keV})^{0.86} \right]^{-1}$~cm$^2$~g$^{-1}$, where $X$ is the hydrogen mass fraction in the photosphere \citep{lewin.1993zr}. Following the assumption that we observed a H-rich burst, we use $X=0.7$ (corresponding to solar composition) and find $L_{Edd}=1.7 \times 10^{38}$ erg~s$^{-1}$. In the time-resolved spectral analysis of the burst, we found that the brightest slice has a bolometric flux $F_{pk}$ =   $3.58^{+0.29}_{-0.28} \times 10^{-8}$ erg~cm$^{-2}$~s$^{-1}$, which leads to a distance of $6.3 \pm 0.5$ kpc. Table \ref{tab:dist} summarizes the distances that have been found in past studies for this source, and shows the dependency of the distance on the Eddington luminosity. The present peak flux is consistent with the brightest bursts that have been observed for this source in the past. Our distance measurement is consistent  with the lower range that had been established by \citet{cocchi.1999fk}, \citet{galloway.2008fk} and \citet{trap.2009vn}. 

The largest source of uncertainty in this calculation is the hydrogen mass fraction $X$ chosen for the NS atmosphere in the Eddington luminosity calculation, where any $X$ value lower than 0.7 would place the source further (Table  \ref{tab:dist}). Uncertainty on the distance also comes from the NS mass; for instance, considering $M=2M_{\odot}$ leads to a distance 9.2\% larger. Given these uncertainties, our distance of $6.3 \pm 0.5$ kpc should be regarded as a lower limit, consistent with the value of $d=7$ kpc that we have been using throughout this paper. Our measurement of the column density towards \GRS ($\sim 10^{23}$ cm$^{-2}$) is consistent with the source being closer than the Galactic center \citep[\nus measured $N_H=1.66^{+0.70}_{-0.61} \times10^{23}$ cm$^{-2}$ to the Galactic center during flares from Sagittarius A$^{\star}$;][]{barriere.2014uq}.

\begin{deluxetable*}{lcccc}
\tabletypesize{\footnotesize}
\tablecaption{Distance for \GRS from this study compared to past work. \label{tab:dist}}
\tablecolumns{5}
\tablewidth{0pt}
\tablehead{
\colhead{Reference} & \colhead{$F_{pk}$ } 		& \colhead{ $d$ (kpc) } 	&  \colhead{ $d$ (kpc) }	& \colhead{ $d$ (kpc) } \\
 				& \colhead{(10$^{-8}$ erg cm$^{-2}$ s$^{-1}$)} 	& \colhead{$L_{\rm Edd} \, (X=0.7)$}		& \colhead{$L_{\rm Edd} \,(X=0)$}	& \colhead{$L_{\rm Edd}$ (Kuulkers) }}
\startdata
\citet{cocchi.1999fk} 		& $3.26 \pm 0.26$ & $^*6.65\pm0.53$ & $^*8.67\pm0.69$ & $9.87\pm0.79$ \\ 
\citet{galloway.2008fk} 	&$3.80 \pm 1.0$ &  $^*6.16\pm1.6$ & $^*8.03\pm2.1$ & $9.14 \pm 2.4$ \\
\citet{trap.2009vn} 		& $5.5^{+0.7}_{-2.1}$ & $<5.12^{+2.0}_{-0.65}$ &  $<6.68^{+2.5}_{-0.85}$ & $^* <7.60^{+2.9}_{-0.97}$ \\
This work 				& $3.58^{+0.29}_{-0.28}$ & $^*6.35^{+0.50}_{-0.51}$ & $8.27^{+0.65}_{-0.66}$ & $9.42^{+0.74}_{-0.75}$ 
\enddata
\tablecomments{The second column shows the burst peak bolometric flux that the different studies used for their distance measurement. The three rightmost columns show the distance they obtained for different values of the Eddington luminosity: $1.7 \times 10^{38}$ erg~s$^{-1}$ for $X=0.7$, $2.9 \times 10^{38}$ erg~s$^{-1}$ for $X=0$, and $(3.79\pm0.15) \times 10^{38}$ erg~s$^{-1}$ as determined empirically by \citet{kuulkers.2003fk}. The stars indicate which value of the Eddington luminosity was favored in each study.}
\end{deluxetable*}

Transient achromatic variability is observed between 48~s and 75~s into the burst. The variability is mostly on top of the exponential decay trend, but we can see at 52 s and 54.5 s that the flux goes below the decaying trend, suggesting that transient obscuration is taking place. It is hard to determine if the interval between 75~s and 95~s belongs to the exponential decay trend, or if it is part of the power-law decay with obscuration (Figure \ref{fig:burstdecay}). The latter option might be more physical, as it is hard to explain why the flux would have suddenly risen by a factor 3--4 while transitioning from exponential to power-law cooling. However, the fact that the two curves do not intersect could be an indication that re-heating of the NS took place during the episode of variability.

\GRS  already displayed such late-time variability, as visible in bursts \#3 and possibly \#4 observed by {\it INTEGRAL} JEM-X \citep[although the features were not discussed; ][]{trap.2009vn}. \citet{int-zand.2011uq} suggest that this variability is caused by clouds of electrons in thermal balance with the radiation field scattering the NS thermal emission (Thomson scattering). The clouds are orbiting the NS and alternatively obscure our line of sight and reflect the NS radiation towards us. The presence of these clouds could be explained by the reorganization of the accretion disk that had been disrupted by the burst. 
To our knowledge, this phenomenon has been reported in only six bursts from other systems, all showing super-expansion, where the photosphere is lifted up to thousands of km off the NS surface \citep{van-paradijs.1990kx,strohmayer.2002vn,molkov.2005ys,int-zand.2005zr,int-zand.2011uq,degenaar.2013fk}. It is interesting that the present burst shows similar behavior, although the photosphere has been merely lifted to a few tens of km. Assuming that the two troughs at 52~s and 54.5~s are the signature of a cloud in a Keplerian orbit of period $P=2.5$~s, their elevation above the NS surface would be $(P^2GM/4\pi^2)^{1/3} \approx 3\times10^3$ km (for $M = 1.4M_{\odot}$), well beyond the photosphere. Thus, if this variability was indeed the sign of a disturbed inner accretion disk, it would have been caused only by the action of the Eddington flux and the resulting X-ray heating \citep[e.g.][]{ballantyne.2005ly}. The fact that late time variability is rare could then be explained by the viewing angle, where only high inclination systems would show this phenomenon. We note however, that \GRS is not an eclipsing or dipping system, which limits the validity of this explanation.

The high dynamic range of this burst allows the observation of a long power-law decay tail, starting around t=110~s after the variability ceased. This is consistent with a pure radiative cooling without any additional production of heat. We find a power-law index of 2.25, which indicates that photons contributed significantly to the heat capacity \citep{int-zand.2014uq}.

Oscillations at 589 Hz had been tentatively associated to \GRS (section \ref{sec:oscillations}). Despite {\it NuSTAR}'s ability to measure oscillation frequency up to $\approx$1 kHz \citep[over short periods of time,][]{Bachetti+14}, we did not detect oscillations in the burst nor in the outburst. However, our upper limits are not constraining, being always above the pulsed fraction measured by \citet{strohmayer.1997kx}. In addition, it has been proposed that oscillations appear mostly in He burning runaways \citep{cumming.2000vn, narayan.2007ys}, which was not the case here, and it is known that they are not detected consistently even in He bursts from sources known to produce pulsed emission. So we cannot confirm nor repudiate the association of the 589 Hz pulsation with \GRS.

The spectrum of the PRE phase shows weak evidence of an absorption line at $5.46 \pm 0.10$ keV (95.6\% confidence level). Although this is only marginal detection, the importance of these features makes it worthwhile to consider its interpretation. One additional motivation for this is that there is a precedent: a line at $5.7 \pm 0.25$~keV was detected in the rise and PRE phase of a bright burst from 4U~1636-536 observed with {\it Tenma} a.k.a. {\it Astro-B} \citep{waki.1984vn}. Although this detection was clear ($> 99.5$\% confidence level), no similar line was ever observed subsequently in any source. 

Detection of absorption lines in burst spectra are rare and have been controversial due to their lack of consistency, even within seemingly identical bursts from given sources. Here is a brief summary of the main cases. A line at 4.1 keV was detected in the {\it Tenma} spectra of non-PRE bursts from 4U 1636-536 \citep[in 4 out of 12 bursts,][]{waki.1984vn} and from 4U 1608-52 \citep[in 3 out of 17 bursts,][]{nakamura.1988ys}. Another line at 4.1 keV was found by \citet{magnier.1989zr} with {\it EXOSAT} in a 0.5~s slice of the rise of a non-PRE burst from  EXO 1747-214, although this detection is more controversial due to its short duration and large equivalent width. More recently, \citet{cottam.2002vn} found a set of absorption lines in the stacked spectra of 28 bursts from EXO~0748--676  observed with the Reflection Grating Spectrometer onboard {\it XMM-Newton}, which they interpreted as being due to transitions from Fe XXV, Fe XXVI,  and O VIII, all gravitationally redshifted by a factor $1+ z = 1.35$. This detection was not confirmed in subsequent observations \citep{cottam.2008uq}, and a 552~Hz spin was found for the NS \citep{galloway.2010uq}, which is incompatible with the production of narrow absorption lines at the surface of a NS \citep{lin.2010fk}.
\citet{int-zand.2010ly} analyzed 32 super-expansion bursts and detected absorption edges in three cases. They interpreted them as signatures of heavy elements (synthesized during the burst) present in the atmosphere of the NS and the wind created by the burst.

If the \GRS line is real, what could its origin be? The two mechanisms that can produce absorption lines are resonant  cyclotron scattering and resonant atomic level transitions. An electron cyclotron line at 5.5 keV would require a magnetic field of the order of 10$^{12}$~G, which is 3--4 orders of magnitude higher than what is expected for Type I bursters \cite[e.g.][]{bildsten.1998fk}. Thus we favor the resonant atomic transition interpretation. 

During the PRE phase of the present burst, we observe a roughly constant color temperature of $T = 2.44 \times 10^{7} K$ ($k_BT = 2.1$ keV), which corresponds to a local effective temperature of  $T_{e\star} \approx 23$ MK, using a correction factor of 1.4 to account for the spectral hardening from Comptonization \citep{madej.2004zr}, and correcting for the redshift with $1+z = 1.31$. We assume that accretion is quenched during the PRE phase of the burst due to the radiation pressure at the Eddington level. \citet{weinberg.2006uq} showed that the products of nuclear burning can come to sufficiently shallow depth to be blown away by the wind and therefore be exposed to the photosphere. According to the same authors, the most likely heavy elements to be present in the photosphere and above it in the wind are Zn and Cr for a mixed H/He burst. At a temperature of $T_{e\star} \approx 23$ MK, Saha equilibrium indicates that these elements will be either fully ionized or in the H-like ionization state. The most intense transition for H-like ions is K$\alpha$\footnote{The K$\alpha$ transition is actually double, K$\alpha_{1}$ referring to $1s \rightarrow 2p_{3/2}$ and K$\alpha_{2}$ referring to $1s \rightarrow 2p_{1/2}$, with K$\alpha_{1}$ twice as strong as K$\alpha_{2}$. However, the energy difference is too small to differentiate with \nus.}. We see in Table \ref{tab:absline} that if Zn was causing this absorption line, the gravitational redshift factor would be $1+z = 1.70$, which is not reasonable during a PRE phase. However, Cr yields a factor $1+z=1.09$, resulting in a plausible radius range of 29.0 -- 41.4 km for a mass range of 1.4 -- 2.0 $M_{\odot}$.

Besides the obvious limitation that the line is only weakly detected, one caveat of this interpretation is that
the accretion rate preceding this burst was lower than $0.01 \dot{m}_{\rm Edd}$. This goes beyond the case studied by \citet{weinberg.2006uq} (accretion rate of $\approx 0.01 \dot{m}_{\rm Edd}$ resulting in a mixed H/He burst triggered by triple-$\alpha$ He reactions), which could result in very different nucleosynthesis products being sent in the NS atmosphere.

\begin{deluxetable}{ccc}
\tabletypesize{\footnotesize}
\tablecaption{K$\alpha_{1} $ transition energy of H-like ions {\it possibly} present in the atmosphere of the NS, and corresponding division factor to shift the line to 5.46 keV \label{tab:absline}}
\tablecolumns{3}
\tablewidth{0pt}
\tablehead{
\colhead{Ion} & \colhead{$E_{{\rm K}\alpha}$ (keV) } 	&   \colhead{ $(1+z)$ }
}
\startdata
Zn	& 9.31		& 1.71 \\
Ni 	& 8.10 	 	& 1.48 \\
Fe 	& 6.97 	 	& 1.28 \\
Mn 	& 6.44 	 	& 1.18 \\
Cr 	& 5.93 	 	& 1.09 \\
V 	& 5.44 	 	& 1.00
\enddata
\end{deluxetable}


\section{Summary}
\nus observed \GRS twice in Summer 2013 as it was entering in outburst for the first time since July 2010. We detected a Type I burst with the second highest fluence reported for this source (burst \#8 in \citet{galloway.2008fk} has 50\% higher fluence). Based on the low accretion rate before the burst ($3.0 \times 10^{-3}$~$\dot{m}_{\rm Edd}$) and the long duration of the burst, which includes a 18-s plateau, we interpret it as an H-triggered mixed H/He thermonuclear explosion. The burst underwent a moderate phase of PRE, thus reaching the Eddington luminosity.

Assuming an H-rich atmosphere, we derive a new constraint on the source distance, $6.3 \pm 0.5$~kpc. However, given the fact that this value relies on highly uncertain parameters, such as the hydrogen mass fraction in the atmosphere, and the NS mass, we consider it more conservative to use the distance of 7~kpc, in agreement with past works.

The low persistent emission allows a power-law decay to be revealed in the burst tail. The transition from the exponential to the power-law decay is affected by strong variability above and below the decay trend. To our knowledge, it is the first time that this phenomenon is reported in a moderate PRE burst. The current model involving clouds of electrons created as the inner accretion disk reorganizes itself after having been disturbed by the burst is hard to reconcile with the fact that this late time variability is rare.

We searched for pulsations during the burst and the outburst, but found none. However our upper limits on the pulsed fraction around 589 Hz (between 17\% and 25\%) during the peak and the decay of the burst are not constraining as they are above the strongest pulsed fraction reported by \citet{strohmayer.1997kx}.

A line at $5.46 \pm 0.10$ keV is weakly detected during the PRE phase of the burst. Assuming that the line is real, we speculate that it is due to the K$\alpha$ resonance absorption line of H-like Cr ions, which is a product of the nuclear reactions, based on the work of \citet{weinberg.2006uq}. Unfortunately, given that the line is observed while the atmosphere is lifted off the surface by radiation pressure, its gravitational redshift carries no information about the NS mass and radius.

\acknowledgments
This work was supported under NASA Contract No. NNG08FD60C, and
made use of data from the {\it NuSTAR} mission, a project led by
the California Institute of Technology, managed by the Jet Propulsion
Laboratory, and funded by the National Aeronautics and Space
Administration. We thank the {\it NuSTAR} Operations, Software and
Calibration teams for support with the execution and analysis of
these observations.  This research has made use of the {\it NuSTAR}
Data Analysis Software (NuSTARDAS) jointly developed by the ASI
Science Data Center (ASDC, Italy) and the California Institute of
Technology (USA). The authors wish to thank Nevin Weinberg for useful 
discussions, and the anonymous referee for constructive comments.\\



{\it Facilities:} \facility{NuSTAR}.



\bibliography{GRS17419}


\clearpage



\end{document}